\pdfoutput=1

\documentclass[11pt]{article}

\usepackage[]{acl}

\usepackage{times}
\usepackage{latexsym}

\usepackage[T1]{fontenc}

\usepackage[utf8]{inputenc}

\usepackage{microtype}

\usepackage{graphicx}
\usepackage{multirow}
\usepackage{latexsym}
\usepackage{amsmath,amsthm}
\usepackage{amssymb}
\usepackage{algorithm}
\usepackage{mathrsfs}
\usepackage[noend]{algpseudocode}
\usepackage{booktabs} 
\usepackage{caption} 
\usepackage{algpseudocode}
\usepackage{pifont}
\usepackage{bm}
\usepackage{comment}

\def\1{{\bf{1}}}
\def\0{{\bf{0}}}

\def\c{{\bf c}}

\def\e{{\bf e}}

\def\h{{\bf h}}

\def\k{{\bf k}}

\def\s{{\bf s}}

\def\w{{\bf w}}
\def\x{{\bf x}}

\def\Dcal{{\mathcal{D}}}

\def\Lcal{{\mathcal{L}}}

\title{Deep Keyphrase Completion}

\author{Yu Zhao\textsuperscript{1,}\thanks{\ \ Corresponding author: Y. Zhao (zhaoyu@swufe.edu.cn).}, Jia Song\textsuperscript{1}, Huali Feng\textsuperscript{1}, Fuzhen Zhuang\textsuperscript{2,3}, Qing Li\textsuperscript{1}, Xiaojie Wang\textsuperscript{4}, Ji Liu\textsuperscript{5}\\
  \textsuperscript{1}Fintech Innovation Center, Financial Intelligence and Financial Engineering Key Laboratory, \\ Southwestern University of Finance and Economics, Chengdu, China \\
 \textsuperscript{2}Institute of Artificial Intelligence, Beihang University, Beijing 100191, China\\
  \textsuperscript{3}Xiamen Data Intelligence Academy of ICT, CAS, China\\
 \textsuperscript{4} School of Artifical Intelligence, Beijing University of Posts and Telecommunications, China.\\
 \textsuperscript{5}AI platform and Seattle AI Lab, Kwai Inc.}

\begin{document}
\maketitle
\begin{abstract}
Keyphrase provides accurate information of document content that is highly compact, concise, full of meanings, and widely used for discourse comprehension, organization, and text retrieval. Though previous studies have made substantial efforts for automated keyphrase extraction and generation, surprisingly, few studies have been made for \textit{keyphrase completion} (KPC). KPC aims to generate more keyphrases for document (e.g. scientific publication) taking advantage of document content along with a very limited number of known keyphrases, which can be applied to improve text indexing system, etc. 
In this paper, we propose a novel KPC method with an encoder-decoder framework. We name it \textit{deep keyphrase completion} (DKPC) since it attempts to capture the deep semantic meaning of the document content together with known keyphrases via a deep learning framework. Specifically, the encoder and the decoder in DKPC play different roles to make full use of the known keyphrases. 
The former considers the keyphrase-guiding factors, which aggregates information of known keyphrases into context. On the contrary, the latter considers the keyphrase-inhibited factor to inhibit semantically repeated keyphrase generation.
Extensive experiments on benchmark datasets demonstrate the efficacy of our proposed model.
\end{abstract}

\section{Introduction}
\begin{figure}[t]
    \centering
    \includegraphics[width=0.5\textwidth]{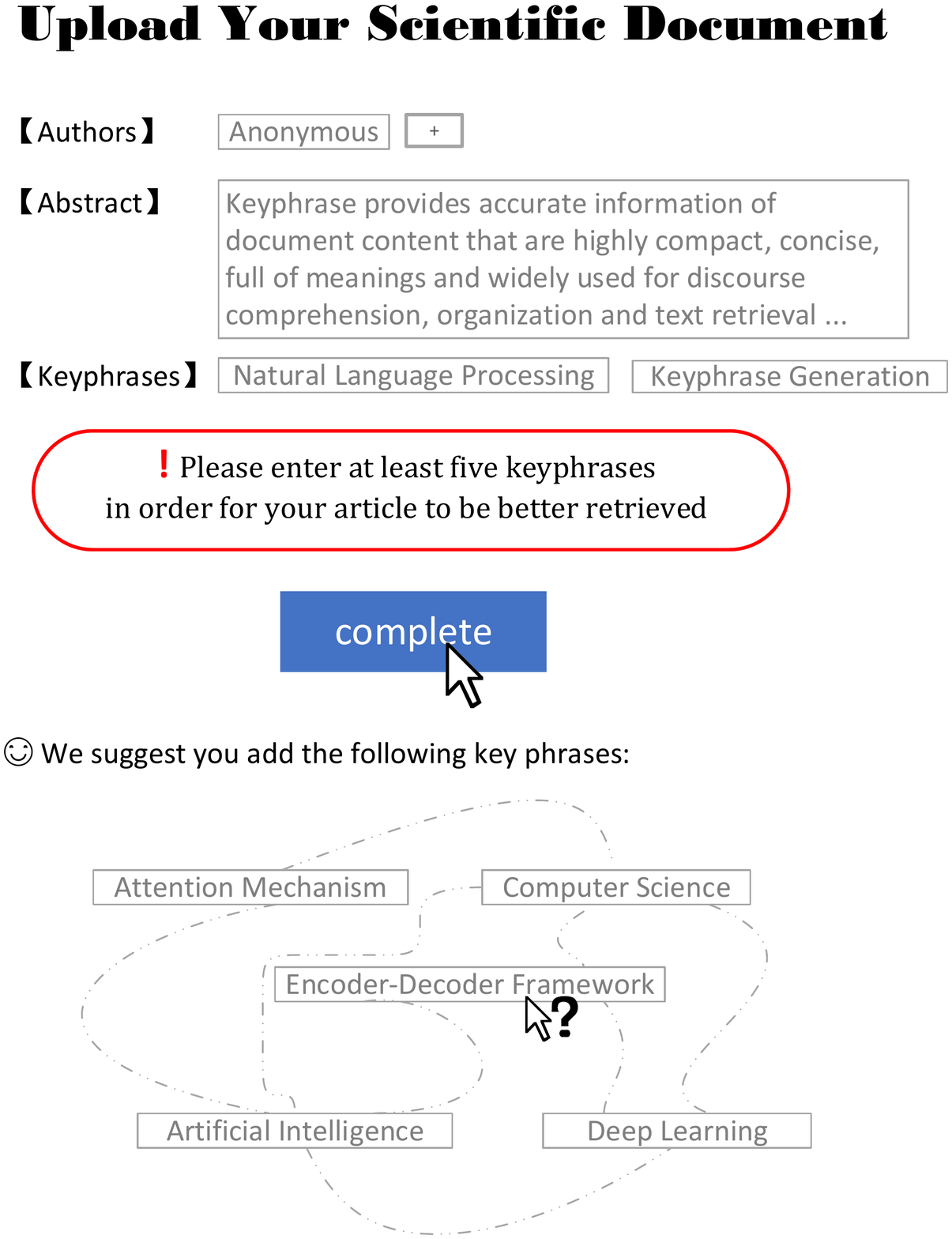}
    \caption{Example of keyphrase completion for text retrieval system.}
    \label{figure-example}
\end{figure}

Keyphrases are highly concise phrases that can provide compact and accurate information of given document content. Since high-quality keyphrases can facilitate the understanding, organizing, and accessing of document content, 
they are wildly used for NLP tasks, such as 
document clustering \cite{Hammouda2005Corephrase,Chiu2020Autoencoding}, 
text summarization \cite{Qazvinian2010Citation,Cano2019Keyphrase},
and text retrieval \cite{Boudin2020Keyphrase}. 
Due to the broadly demand, many automatic keyphrases extraction and generation methods are proposed over the years \cite{Meng2017Deep}. 



In this paper, we concentrate on the problem of \textbf{\textit{keyphrase completion} (KPC)}, which aims to predict a relative large fixed size of keyphrases for document management. 
Consider the following situation where a publications retrieval system is going to be established and improved, each academic paper is demanded for \textit{N} keyphrases to fill in while it only provides a very limited number of known ones (i.e. much lower than \textit{N}) and cannot meet the demand. In this kind of cases, it needs and expects for KPC, i.e. to predict more keyphrases for completion.

Most existing keyphrase prediction models, whether they be extracted-based methods \cite{Mihalcea2004Textrank,Wan2008Single} or generated-enhanced methods \cite{Meng2017Deep}, solely take advantage of the content of document. In fact, as we mentioned before, a document like scientific publication usually has already provided a few known keyphrases, with indicative topic information for the paper. 
For example, in Figure \ref{figure-example} we show an example of keyphrase completion for text retrieval system, in which the scientific document have already existed two known keyphrases and be demanded at least five ones to complete. 
A serious drawback of using existing models for KPC is that they ignore the role of the known keyphrases and consequently fail to consider the already summarized information about them.






In this paper, we propose a novel KPC method with an encoder-decoder framework. We name it as \textit{Deep KeyPhrase Completion} (DKPC) since it attempts to capture the deep semantic meaning of the document content together with known keyphrases via a deep learning framework. 


Specifically, we believe that the known keyphrases have different roles in keyphrase completion, including keyphrase-guiding factors and keyphrase-inhibited factors. Firstly, we take the keyphrase-guiding factors into consideration for the encoder, which aggregates information of known keyphrases into context.
Secondly, we take the keyphrase-inhibited factor into consideration for the decoder, which inhibits semantically repeated keyphrase generation. 
We conduct a comprehensive comparison on four benchmark datasets against two unsupervised models and two supervised deep learning models as baselines, and the results demonstrate the effectiveness of our proposed model.

The {contributions} of this paper are threefold:
\begin{itemize}
    \item To the best of our knowledge, this is the first work that considers the problem of keyphrase completion using the deep learning method.
    \item We propose a novel KPC method for keyphrase completion using an encoder-decoder framework.
    \item We conduct a comprehensive comparison on four benchmark datasets against four baselines, and the results demonstrate the effectiveness of our proposed model.
\end{itemize}




\section{Related Work}
\label{related-work}

\subsection{Keyphrase Extraction}
Extractive models aim to extract present keyphrases from text documents, which can be broadly categorized into two approaches: \textit{unsupervised} and \textit{supervised} \cite{Kim2013Automatic,Hu2006Automatic,Nguyen2007Keyphrase,Augenstein2017SemEval}. 

\subsubsection{Unsupervised Approaches.} 
They are mostly based on statistical models like TF-IDF, clustering and graph-based ranking. 
Generally, they directly treat keyphrase extraction as a ranking task. 
For example, inspired by PageRank \cite{Brin1998The}, \citet{Mihalcea2004Textrank} propose TextRank, which is the first attempt to utilize an unsupervised graph-based approach to rank keyphrases on a word graph.
Following this framework, a substantial amount of enhanced unsupervised keyphrase extraction models are proposed via exploring various word graph from different perspectives. 
For example, 
SemanticRank \cite{Tsatsaronis2010SemanticRank}
take account of semantic relatedness on word graphs for extracting keyphrases. 
CiteTextRank \cite{Gollapalli2004Extracting} and ExpandRank \cite{Wan2008Single} use neighborhood knowledge, such as citation networks \cite{Gollapalli2004Extracting} and topic-related documents \cite{Wan2008Single}, to improve keyphrase extraction. 
Moreover, 
MIKE \cite{Zhang2017MIKE} integrates multidimensional heterogeneous information into a unified framework to improve keyphrase extraction. 
Recently, 
YAKE! \cite{Campos2018YAKE} focuses on {multi-lingual} keyword extraction from single documents. 
{Key2vec} \cite{Mahata2018Key2Vec} leverages {phrase embedding} for keyword extraction in scientific articles. 

\subsubsection{Supervised approaches.} They typically treat keyphrase extraction problem as a binary classification task (i.e., \textit{keyphrase},$\lnot$\textit{keyphrase}) \cite{Hulth2003Improved}, where a classification model is trained on various features 
(i.e., \textit{tf.idf}, candidate length, POS tags, section information, frequency) 
of labelled keyphrases to determine candidate phrase. 
Current supervised keyphrase extraction algorithms includes 
Decision Trees \cite{Sterckx2016Supervised}, 
Support Vector Machine (SVM) \cite{Lopez2010HUMB},
Graph Attention Networks (GAT) \cite{Prasad2019Glocal}. 
For example, 
\citet{Caragea2014Citation} proposed to take citation network information into consideration for keyphrase extraction. 
\citet{Prasad2019Glocal} incorporated word importance in graph attention networks for keyphrase extraction. 
Recently, some methods treat keyphrase extraction as {sequence labeling task} \cite{Gollapalli2017Incorporating,Zhang2016Keyphrase,Chowdhury2019Keyphrase,Al-Zaidy2019Bi-LSTM-CRF}.
For instance, 
\citet{Gollapalli2017Incorporating} extracted keyphrases from research papers utilizing token-based features incorporating expert knowledge through sequence labeling. 
\citet{Sun2019Divgraphpointer} proposed an end-to-end method to extract diverse keyphrase by combining traditional-based ranking methods and RNN-based approaches. 
\citet{Al-Zaidy2019Bi-LSTM-CRF} and propose a joint model by combining CRF and BiLSTM.

\subsection{\textbf{Keyphrase Generation}}
Different from keyphrase extraction, keyphrase generation models could generate absent keyphrases by modeling such task as a sequence-to-sequence (seq2seq) learning problem.  
\citet{Meng2017Deep} first proposed CopyRNN, a seq2seq framework with copy mechanism for keyphrase generation. 
Following this framework, many enhanced variants of CopyRNN are proposed \cite{Ye2018Semi-supervised,Chen2019Title-Guided,Cano2019Keyphrase,Chan2019Neural,Yuan2020One,Chen2020Exclusive,Liu2020Keyphrase}. 
For example, 
\citet{Ye2018Semi-supervised} proposed a semi-supervised methods by leveraging both labeled and unlabeled data. 
\citet{Chen2019Title-Guided} proposed TG-Net taking title information into consideration. 
\citet{Chan2019Neural} utilized reinforcement learning with adaptive rewards to generate more sufficient and accurate keyphrase. 
\citet{Chen2020Exclusive} designed a hierarchical decoding process and an exclusion mechanism to avoid generating duplicated keyphrases. 
Moreover, \citet{Swaminathan2020Keyphrase,Swaminathan2020A} proposed a keyphrase generation approach using conditional Generative Adversarial Networks. 
\cite{Ye2021One2Set} proposed a new training paradigm ONE2SET without predefining an order to concatenate the keyphrases. 
\cite{Ahmad2021Select} proposed a method for neural keyphrase generation with layer-wise coverage attention.

\section{Methodology}
In this section, we introduce the details of our method.
As we stated before, the existing keyphrases play two kinds of roles in keyphrase completion, as follows:
\begin{itemize}
    \item \textit{\textbf{Guiding Factor}} Since keyphrases are a group of words that condense the core information of a document, they always belong to the related topics. For instance, the keyphrases of this paper, ``keyphrase generation" and "attention mechanism" are both related to Natural Language Processing. Motivated by this observation, we can capture such relevance and let existing ground-truth keyphrases to guide us complete more keyphrases. We call it the guiding factor.
    \item \textit{\textbf{Inhibiting Factor}} It is reported that the keyphrases of more than 85\% documents in the largest keyphrase generation benchmark dataset have different first word \cite{Chen2020Exclusive}. In the generation stage, if the first word of the predicted keyphrase is the same as one of the existing keyphrases, it will be eliminated. We call it inhibiting factor.
\end{itemize}
Therefore, motivated the basic ideas above, our proposed model contains two modules: keyphrase-guided encoder and keyphrase-inhibited decoder. Before we introduce our model, we first give a formal problem definition.

\subsection{Problem Definition}
The task of keyphrase generation is usually formulated as follows: given a text dataset $\Dcal{} = \{{\bf x}^i, {\bf y}^i\}_{i=1}^N$ where ${\bf x}_i$ is $i$-th source text and ${\bf y}^i=\{{\bf y}^{i,1}, {\bf y}^{i,2},...,{\bf y}^{i, M}\}$ is a set of keyphrases of it. $N$ is the number of documents, and $M$ is the number of keyphrases of the $i$-th document.
Both the ${\bf x}_i$ and the $j$-th keyphrase of it are sequences of words, denoted as ${\bf x}^i=(^i_1,x^i_2,...,x^i_{L_{x^i}})$ and ${\bf y}^{i,j}=(y_1^{i,j},y_2^{i,j},...,y_{L_{y^{i,j}}}^{i,j})$, where $L_{x^i}$ and $L_{y^{i,j}}$ are the number of words of the i-th texts and its j-th keyphrase respectively.

The goal of the model is to map from ${\bf x}$ to ${\bf y}$. Since our task is complete more keyphrases based on document content and a limited number of known keyphrases, we split the data into $({\bf x}, {\bf k}, {\bf y})$ where $\bf x$ is source text, ${\bf k}$ is the given existing keyphrases of document $x$ in advance, and ${\bf y}$ is the rest true keyphrases that we expect to generate for document completion.

\subsection{Keyphrase-Guided Encoder}
In the keyphrase-guided encoder, we aim to utilize the known keyphrases to guide more keyphrase generation. To this end, we take advantage of the promising attention mechanism in the encoder, which can gather the information of given keyphrases to the source text.

\paragraph{Source Text Representation.}
For words in source text, we use a bi-directional GRU \citet{cho-etal-2014-learning} to learn the contextual representation. The word embeddings are obtained from an initial embedding lookup table. 
The encoder Bi-GRU reads the input sequence $\bf x=(x_1, x_2,..., x_{L_x})$ forwardly and backwardly to converts them into two sets of hidden representation by iterating with the following equation along time $t$, respectively. 

\begin{equation}
\overrightarrow{\h^x_t}=f(\x_t,\overrightarrow{\h^x_{t-1}}) \ ,
\end{equation}

\begin{equation}
\overleftarrow{\h^x_t}=f(\x_t,\overleftarrow{\h^x_{t-1}}) \ ,
\end{equation}
where $f$ is a non-linear function. $\overrightarrow{\h_t^x}$ is the hidden state at time $t$ when input sequence is ordered from $\x_1$ to $\x_{L_x}$ and $\overleftarrow{\h_t^x}$ is the same when input sequence is ordered from $\x_{L_x}$ to $\x_{1}$. $L_x$ is the number of words of source text $\x$. 
The forward hidden state and backward hidden state are concatenated as $\h^x_i=[\overrightarrow{\h_i^x};\overleftarrow{\h_i^x}]$ to represent the $i$-th word of $\x$.

\paragraph{Keyphrase Representation.}
Simple averaging is sufficient for keyphrase representation since keyphrase typically consists of a small number of words (in most cases 1 or 2). On the contrary, complicated models (like CNN or RNN) tend to overfit \cite{Xin2018Improving}. 
Therefore, similar to \cite{Xin2018Improving}, if a given keyphrase contains $n_k$ words, we use the average of embedding vectors of the $n_k$ words to represent this keyphrase, as follows:

\begin{equation}
    \k = \frac{\w_1+\w_2+...+\w_{n_k}}{n_k} \ ,
\end{equation}
where $\k$ is an aggregated semantic representation vector and $w_i$ is the embedding vector of $i$-th word of this keyphrase from the embedding lookup table. In the phrase of model training, these vectors will be constantly updated to learn more appropriate semantic representations.

\paragraph{Attentional Hidden State.} 
To effectively exploit the information provided by given keyphrases, an attention layer is used to calculate the relevance of given keyphrases and source text, and then obtain an aggregated information vector. This formal process is as follows:
\begin{equation}
    e_{ij}=(\h_i^x)^\top\bf{W_1} {\k_j} \ ,
\end{equation}

\begin{equation}
    \alpha_{i,j}=\frac{exp(e_{ij})}{\sum^{n_K}_{n=1}exp(e_{in})} \ ,
\end{equation}

\begin{equation}
    \c_i=\sum^{n_k}_{j=1}\alpha_{i,j}\k_j \ ,
\end{equation}
where  $\alpha_{ij}$ is the normalized attention score of the i-th word of source text and the j-th given keyphrase. $\c$ is an aggregated information vector which summarizes the information of all the given keyphrases. 
It is calculated based on the similarity of two vectors. We use the general mode as our alignment function \citet{luong-etal-2015-effective}.

Finally, we concatenate the aggregated information vector of given keyphrases $\c$ and the source text representation $\h^x$ as encoder hidden state $\h$ for the downstream decoding process. 
\begin{equation}
    \h=\ [\c \| \h^x \ ]
\end{equation}
Therefore, we can obtain $[\h_1,\h_2,...,\h_{L_x}]$ according to this formula at different time steps. With the proposed keyphrase-guided encoder, the existing keyphrases can provide some useful information related to the domain of the source text and then guide to complete more ground-truth keyphrases.

\subsection{Keyphrase-Inhibited Decoder}
In the keyphrase-inhibited decoder, we aim to inhibit semantically repeated keyphrase generation. To this end, we employ Bi-GRU with attention mechanism and copy mechanism \cite{gu-etal-2016-incorporating} as the building blocks of the decoder. 
It decompresses the source text into a context vector through an attention layer and generates the target keyphrase word by word. The process of Decoder is designed as follows:

\begin{equation}
 \s_t={\text{GRU}}([\e_{t-1};\zeta_{t-1}], \s_{t-1}) \ ,
\end{equation}

\begin{equation}
    \c_t=attn(\s_t,[\h_1,\h_2,...,\h_{L_x}],\bf{W_2}) \ ,
\end{equation}

\begin{equation}
 \widetilde{\h_t}=tanh(\bf{W_3}[\c_t;\s_t]) \ ,
\end{equation}
where $\e_{t-1}$ is the embedding vector of $y_{t-1}$ and $\zeta_{t-1}$ is position representation of it by selective reading in copy mechanism. The context vector $\c_t$ is the aggregated vector for $\s_t$ from the encoder's source text representation $[\h_1,\h_2,...,\h_{L_x]}$ via attention mechanism.

Target hidden state $\s_t$ and the context vector $\c_t$ that weighs the sum of the hidden state of the encoder are combined through a simple concatenation layer to produce an attentional hidden sate. Later, the attentional hidden state goes through a softmax layer and outputs the probability distribution. Note that the probability of a word be produced is composed of two parts: the probability of generating it and the probability of coping it from source text. The process is as follows:

\begin{align}
    p(y_t|y_{t-1},\mathbf{x},\mathbf{k}) = & p(y_t,g|y_{t-1},\mathbf{x},\mathbf{k})+\\ &p(y_t,c|y_{t-1},\mathbf{x},\mathbf{k}) \ ,
\end{align}
    
where $p(y_t,c|·)$ is the probability of $y_t$ being copied from source text and $p(y_t,g|·)$ is the probability of $y_t$ being generating from the large vocab. Specifically, the calculation process is as follows:

\begin{equation}    
    p(y_t,g|·)=softmax(\mathbf{W_4}\widetilde{\h_t})\ ,
\end{equation}
\begin{equation}  
    p(y_t,c|·)=tanh(\mathbf{W_5}\h_t)\s_t \ .
\end{equation}

Finally, normalize the sum of the two probabilities as the final probability that $y_t$ is generated at the current step. If 
$y_t$ is the same as the initial word of one of the given keyphrases, then it won't be the output as result.

\subsection{Training}

The widely used negative log likelihood loss is adopted to train our model:

\begin{equation}    
    \Lcal{} = -\sum^n_{t=1}\log p(y_{t}|\mathbf{y_{t-1}},\mathbf{x},\mathbf{k}) \ ,
\end{equation}
where $\mathbf{y}_t$ is the t-th predicted word sequence. $n$ is the length of target keyphrase $\mathbf{y}_t$, $y_t$ is the t-th word of which.

\section{Experiments}
\label{experiments}
\subsection{Experimental Settings} 


\begin{table*}[t]\footnotesize
    \centering
    \caption{The performance of expanding keyphrases of various models on four benchmark datasets.}
    \label{table:my_label}
    \newcommand{\tabincell}[2]{\begin{tabular}{@{}#1@{}}#2\end{tabular}}
    \centering
    \begin{tabular}{l||cc|cc|cc|cc}
    \toprule    
    \multirow{2}*{\bf{Methods} } &\multicolumn{2}{c|}{\bf Inspec} &\multicolumn{2}{c|}{\bf Krapivin}&\multicolumn{2}{c|}{\bf SemEval}&\multicolumn{2}{c}{\bf KP20k}\\
       &\textbf{F1@5}&\textbf{F1@10}&\textbf{F1@5} &\textbf{F1@10}&\textbf{F1@5} &\textbf{F1@10}&\textbf{F1@5} &\textbf{F1@10} \\
    \midrule
    \midrule
    TF-IDF &0.107   &0.151  &0.068  &0.067   &0.061   &0.073 &0.108   &0.101  \\
 
    TextRank &0.172   &0.190  &0.076  &0.084  &0.028   &0.039 &0.065   &0.065  \\
    
    \midrule
    RNN   &0.084   &0.082  &0.068  &0.060  &0.077   &0.084 &0.103   &0.087   \\

    CopyRNN &0.179   &0.194  &0.127  &0.106  &0.106   &0.128 &0.161   &0.142 \\
    \midrule
    \textbf{DKPC (Ours)}&\textbf{0.212}   &\textbf{0.218}  &\textbf{0.161}  &\textbf{0.138}  &\textbf{0.120}   &\textbf{0.141} &\textbf{0.181}   &\textbf{0.154} \\
    \bottomrule
    \end{tabular}
\end{table*}

\subsubsection{Datasets} 
We utilize the public benchmark datasets for the experiments, includes: \textit{KP20k} \cite{Meng2017Deep}, \textit{Inspec} \cite{Hulth2003Improved}, \textit{Krapivin} \cite{2009Large}, \textit{SemEval 2010} \cite{Kim2010SemEval}.

\begin{itemize}
    \item \textbf{\textit{KP20k}} \cite{Meng2017Deep} is the largest dataset on Keyphrase Generation. \textit{KP20k} contains 567,830 scientific articles, where 20,000 articles for training and 20,000 articles for testing. 
    
    \item \textbf{\textit{Inspec}} \cite{Hulth2003Improved} is composed of 2000 abstracts of scientific articles totally. The testing part which is composed of 500 abstracts is used to test our model.

    \item \textbf{\textit{Krapivin}} \cite{2009Large} is not split to testing part and training part by its original authors. We thus follow the set in \cite{Meng2017Deep} that uses the first 400 papers in alphabetical order as the testing part in total 2,304 papers. 


    \item \textbf{\textit{SemEval 2010}} \cite{Mahata2018Key2Vec} consists of 288 full length ACM articles. The testing part has 100 articles. 
\end{itemize}
Note that our model is only trained once using the training of \textit{KP20k}, and evaluated on the testing part of all other datasets, which is a normal strategy like in CopyRNN \cite{Meng2017Deep}, TG-Net \cite{Chen2019Title-Guided}, and ParaNet \cite{Zhao2019Incorporating}. 

\subsubsection{Baselines}
To demonstrate the effectiveness of our proposed model for keyphrase extraction, we compare it with the following baselines, including two unsupervised graph-based methods and two state-of-the-art (SOTA) deep learning models.
\paragraph{Unsupervised graph-based extraction methods:}
\begin{itemize}
    \item TF-IDF: a widely used statistical weighting methods.
    \item TextRank \cite{Mihalcea2004Textrank}: a PageRank based unsupervised algorithm for keyphrase extraction.
\end{itemize}

\paragraph{Deep learning-based generation methods:}
\begin{itemize}
    \item RNN \cite{Meng2017Deep}: a basic model for sequence data, without copy mechanism compared to CopyRNN.
    \item CopyRNN \cite{Meng2017Deep}: a state-of-the-art model which is the first applies the encoder-decoder framework to keyphrase generation.
\end{itemize}

\subsubsection{Evaluation Protocol}
We follow the common practice and evaluate the performance of our model in previous work \cite{Meng2017Deep}. 
Similar to them, we utilize F-measure (F1)@N as an evaluation metric based on the macro-averaged Precision, Recall. 
$N$ indicates the number we need to complete, which is set among \{5, 10 or 50\}. The precision is computed by the number of keyphrases generated correctly over the number of all generated keyphrases. The recall is calculated by the number of keyphrases generated correctly over the number of keyphrases that are the targets of the sample. 
For each method, we give the F-measure at top 5 and top 10 predictions on four real-world benchmark datasets.

\subsubsection{Implementation Details}
During data preprocessing, we tokenize, lowercase, and stemming the text. After that, we use the symbol $<digit>$ to replace each digit. 
Since the task is to use some pre-designated keyphrases and document content to complete more useful keyphrases, we need to take out a few keyphrases before training. 
$M$ denotes the number of original targets of a data record and $N$ denotes the number of pre-designated keyphrases we need to take out at random. We design the processing rules as follows:

\begin{equation}
N = \begin{cases}
delete\ this\ sample, &if\ M =0 \\
1, & if\ M \in \{2,3\} \\
2, & if\ M \in \{4,5\}\\
3, & if\ M \in [6,20] \\
5, & if\ M \textgreater 20 \\
\end{cases}
\end{equation}

The vocabulary includes 50,000 words that appear most frequently in the datasets. 
We set the embedding dimension to 200 and the hidden size to 100. 
We use teacher forcing to help training model. 
One batch contains 128 data samples. 
Our model is optimized by Adam \cite{2014Adam} with an initial learning rate 0.001. 
The learning rate will update every 10000 batches according to StepLR\footnote{A method for interval adjustment of learning rate in PyTorch. }. 
When F1@10 on the validation set doesn't rise for 100 consecutive steps, we stop training early.
During testing phrase, we use beam search to select effective keyphrases and beam size is set to 50.

\begin{table*}[htb]\footnotesize
    \centering
    \caption{The performance of our proposed model using different number of known keyphrases.}
    \label{table-as}
    \newcommand{\tabincell}[2]{\begin{tabular}{@{}#1@{}}#2\end{tabular}}
    \centering
    \begin{tabular}{l||cc|cc|cc|cc|cc}
    \toprule    
    \multirow{2}*{\bf{Methods} } &\multicolumn{2}{c|}{\bf 10\%} &\multicolumn{2}{c|}{\bf 20\%}&\multicolumn{2}{c|}{\bf 30\%}&\multicolumn{2}{c|}{\bf 40\%}&\multicolumn{2}{c}{\bf 50\%}\\
       &\textbf{F1@5}&\textbf{F1@10}&\textbf{F1@5} &\textbf{F1@10}&\textbf{F1@5} &\textbf{F1@10}&\textbf{F1@5} &\textbf{F1@10}&\textbf{F1@5} &\textbf{F1@10} \\
    \midrule
        \midrule
        Inspec &0.182   &0.231 &\textbf{0.212}  &\textbf{0.238}  &0.206   &0.228 &0.190   &0.202 &0.183   &0.187  \\
    \midrule
        Krapivin &0.118   &\textbf{0.200}  &\textbf{0.179}  &0.181  &0.172   &0.167 &0.165   &0.148 &0.147   &0.122  \\
    \midrule
    SemEval &\textbf{0.124}   &\textbf{0.157}  &0.122  &0.142  &0.113  &0.134 &0.098   &0.127 &0.103   &0.122 \\
    \midrule
    KP20k&\textbf{0.243}   &\textbf{0.348}  &0.210  &0.207  &0.205   &0.191 &0.187   &0.166 &0.164   &0.142 \\
    \bottomrule
    \end{tabular}
\end{table*}

\begin{table*}[h]\footnotesize
    \centering
    \caption{The performance of CopyRNN and our proposed model in the same case.}
    \label{table-case}
    \newcommand{\tabincell}[2]{\begin{tabular}{@{}#1@{}}#2\end{tabular}}
    \centering
    \begin{tabular}{l||cc|cc|cc|cc|cc}
    \toprule    
    \bf{Methods}
       &\textbf{F1@5}&\textbf{F1@10}&\textbf{P@5} &\textbf{P@10}&\textbf{R@5} &\textbf{R@10}&\textbf{Similarity}  \\
    \midrule
        \midrule
        CopyRNN &0.400   &0.500 &0.600  &0.500  &0.300 &0.50 &0.387     \\
    \midrule
        DKPC &\textbf{0.533}   &\textbf{0.800}  &\bf 0.800  &\bf 0.800  &\bf 0.400 &\bf 0.800 &\bf 0.473   \\
    \bottomrule
    \end{tabular}
\end{table*}

\subsection{{Results and Analysis}}
Table \ref{table:my_label} shows the performance of our method against four baselines, from which we observe that our proposed model performs better than all baselines for keyphrase completion in terms of both F1@5 and F1@10 metrics on four datasets. The best scores are highlighted in bold. 
It confirms the capability of our method in modeling the deep semantic meaning of document content and known keyphrases for keyphrases completion.

\paragraph{Analysis.} 
(1) To make a fair comparison, we take out a few keyphrases in advance as known keyphrases according to our designed rules mentioned before, which is applied to all testing datasets. In this case, fewer predictable keyphrases will aggravate the keyphrase prediction challenges for all methods, which leads to lower scores\footnote{The absolute scores in Table \ref{table:my_label} are thus lower than their performance in original papers.}. 

(2) As mentioned before, the key motivation behind our work is that we are interested in the proposed model's capability for completing keyphrases based on document content and limited known keyphrases. It is worth noting that such a completion task is a challenging problem. To the best of our knowledge, no existing pertinent methods are proposed specifically to handle this new task. The unsupervised keyphrase extraction models, i.e. TF-IDF and TextRank, and the keyphrase generation models, RNN and CopyRNN, inevitably fail to take the known keyphrases into consideration. We can observe from Table \ref{table:my_label} that all baseline perform worse than ours.

(3) In KPC task, the known keyphrases can play an important role. We believe they are always highly semantically related to other unknown keyphrases that needed to be completed. Our method can capture such relation and utilize known semantic information to select the most conducive sentences to the keyphrases completion from the source document with an attention mechanism. 
In addition, the given keyphrases are also aggregated into the context representation which more fully exploits the information already known. On the four datasets, our method achieves a cumulative 11.3\% improvement at F1@5 and F1@10. 

\subsection{Keyphrase Completion Analysis}
Table \ref{table-as} shows the performance of our proposed model using different numbers of known keyphrases. 
We randomly select 10 percent, 20 percent, 30 percent, 40 percent, and 50 percent of target keyphrases as our known keyphrases in testing data. In each case, we calculate the keyphrase completion performance on all datasets. 
Note that when we extract known keyphrases, we round down the number of original targets.  Therefore, the samples with less than 10 original targets will not be considered in the case when 10 percent keyphrases are extracted. 
Similarly, samples with less than 2 original targets will not be taken into account in the case when 50 percent keyphrases are extracted. 
All the samples that do not meet the minimum keyphrase requirement are deleted.
We can observe from Table \ref{table-as} that the more given keyphrases will not always lead to better model performance. 
The usage of 10 percent or 20 percent keyphrases for completion broadly performs best. 
In KPC task, the appropriate number of known keyphrases is important for model performance. Using too many known keyphrases is not always suitable for predicting the rest of the keyphrase. Because it may generate semantic repeated keyphrases based on the existing keyphrases, therefore, these keyphrases are not able to comprehensively summarize the article.

\begin{figure}[h]
    \centering
    \includegraphics[width=0.5\textwidth]{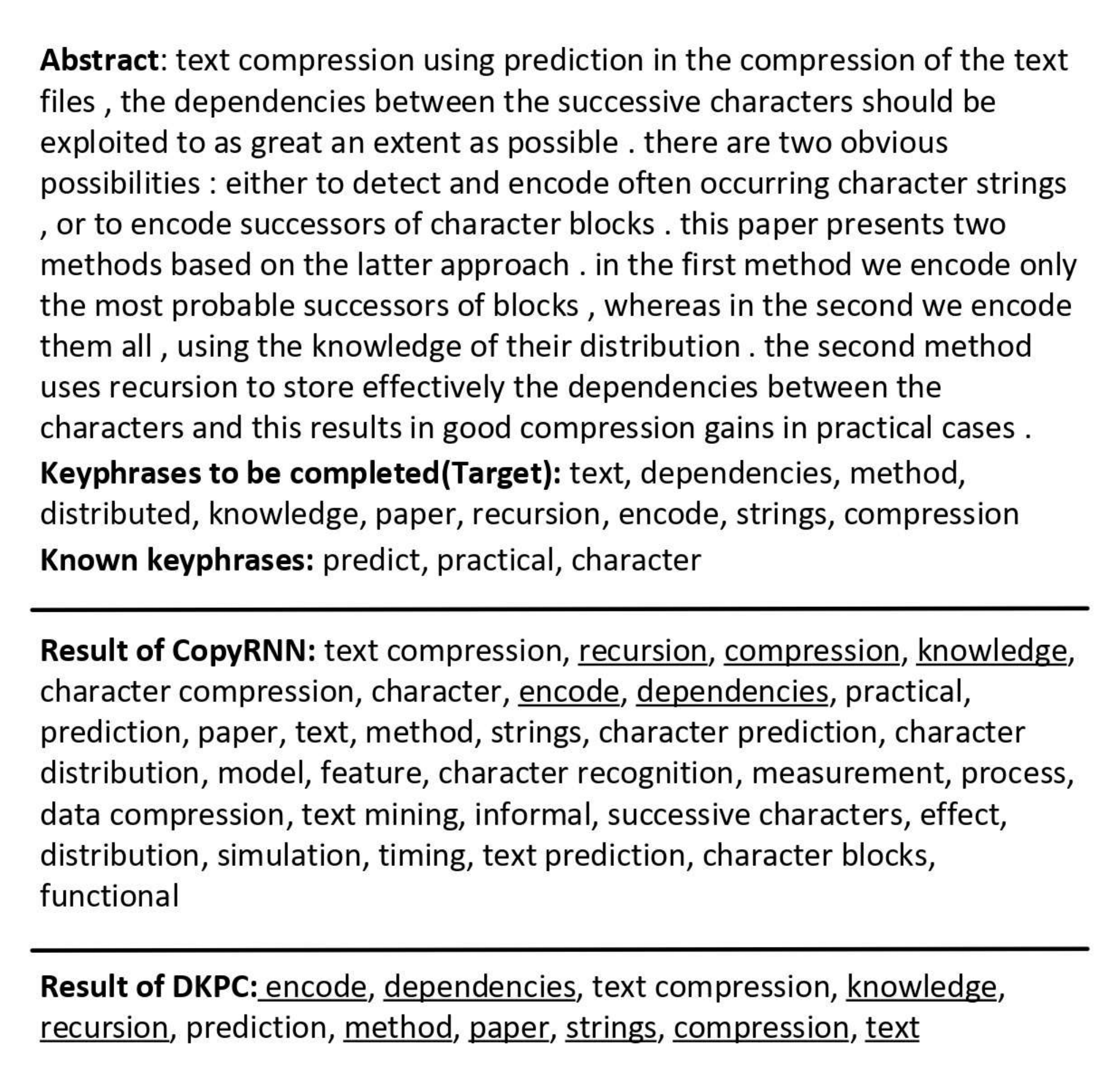}
    \caption{A keyphrase completion example of our proposed DKPC compare to previous method CopyRNN.}
    \label{case_study}
\end{figure}

\subsection{Case Study}
Figure \ref{case_study} and Table \ref{table-case} show a case study of keyphrase completion by our method against CopyRNN. 
we can observe from Figure \ref{case_study} that our model not only completes keyphrases accurately but also completes more targets with less computational resources. 
Table \ref{table-case} shows the performance of the model on Precision, Recall and F1-measure, from which we observe that DKPC is significantly better than CopyRNN. 
Specifically, we use Bert \cite{devlin2019bert} to obtain the embedding of each keyphrase, and then calculate similarity by non-repeating pairs between keyphrases. 
The sum of similarity divided by the number of keyphrase pairs is similarity. 
Due to the guidance of known keyphrases, our proposed method can search for accurate keyphrases from a smaller semantic space than CopyRNN's. That's why our model can find more real keyphrases in less time. The keyphrases inhibition factor is also helpful. For example, in contrast to CopyRNN, our model never includes the known keyphrase ``practical'' as a target to complete, or any keyphrase beginning with the first word of any known keyphrase.

\section{Conclusion and Future Work}
\label{conclusion}
In this paper, we concentrate on the new keyphrase completion task. To the best of our knowledge, this is the first work that considers the problem of keyphrase completion. To solve this problem, we propose a novel KPC method that takes an encoder-decoder framework. The encoder and the decoder in our method make full use of the guidance factor and inhibition factor of known keyphrases, respectively. The comprehensive experimental results demonstrate the effectiveness of our proposed model in the task of keyphrase completion. 
Some interesting future work may include: (1) Our model only explore post-processing for inhibition of known keyphrases. One interesting direction is to design some specific layers on the deep learning model structure to make the inhibition of known keyphrases work better. (2) We use the most classical model, RNN based encoder-decoder framework with attention mechanism and copy mechanism. In the future, we can investigate the performance of other models such as the pre-training model. Some pre-training models trained on the larger dataset may be able to capture the semantic similarity between known keyphrases and keyphrases needed completed better.

\section*{Acknowledgements}
This work was supported by the National Natural Science Foundation of China under Grant No. 61906159, U1836206, 61773361, 62072379, and in part supported by the Fundamental Research Funds for the Central Universities under Grant No. kjcx20210102.

\bibliography{custom}
\bibliographystyle{acl_natbib}




\end{document}